\documentclass[prx,twocolumn,aps,epsf,showpacs,superscriptaddress,longbibliography,nofootinbib]{revtex4-1}
\usepackage[pdftex]{graphicx}

\usepackage{dcolumn}
\usepackage{bm}
\usepackage{epsfig}
\usepackage{latexsym} 
\usepackage{amsmath}
\usepackage{amssymb}
\usepackage{color}
\usepackage{array}
\usepackage{bbm}
\usepackage{hyperref}
\usepackage{color}
\usepackage{array}
\usepackage{cancel}
\usepackage[dvipsnames]{xcolor}
\usepackage{float}

\usepackage{titlesec}
\titleformat{\paragraph}[runin]% runin puts it in the same paragraph
        {\bfseries}% formatting commands to apply to the whole heading
        {}% the label and number
        {0.0em}% space between label/number and subsection title
        {}% formatting commands applied just to subsection title
        [ -- ~]% punctuation or other commands following subsection title
\titlespacing*{\paragraph}{0pt}{4pt}{0pt}

\usepackage[normalem]{ulem} %normalem makes \emph italic rather than underlined

\usepackage{accents}
\newlength{\dhatheight}

\newcommand{\<}{\langle}
\newcommand{\e}{\varepsilon}
\newcommand{\up}{\uparrow}
\newcommand{\down}{\downarrow}
\renewcommand{\>}{\rangle}
\renewcommand{\(}{\left(}
\renewcommand{\)}{\right)}
\renewcommand{\[}{\left[}
\renewcommand{\]}{\right]}
\renewcommand{\d}{\partial}

\newcommand{\eps}{\epsilon}

\newcommand{\Z}{\mathbb{Z}}

\newcommand{\E}{\mathbb{E}}
\newcommand{\tr}{\text{tr}~}

\usepackage{pdfpages} % include appendix
\makeatletter
\AtBeginDocument{\let\LS@rot\@undefined}
\makeatother

\begin{document}
\title{%Volume law entangled phases in monitored random circuits with symmetries\\
%Field-theory of measurement induced phase transitions in symmetric MRCs\\
%Field theory of measurement-induced charge sharpening transitions}
Field theory of charge sharpening in symmetric monitored quantum circuits}
\author{Fergus Barratt}
\affiliation{Department of Physics, University of Massachusetts, Amherst, MA 01003, USA}
\author{Utkarsh Agrawal}
\affiliation{Department of Physics, University of Massachusetts, Amherst, MA 01003, USA}
\author{Sarang Gopalakrishnan}
\affiliation{Department of Physics, The Pennsylvania State University, University Park, PA 16802, USA}
\author{David A. Huse}
\affiliation{Department of Physics, Princeton University, Princeton NJ 08544, USA}
\affiliation{Institute for Advanced Study, Princeton NJ 08540, USA}
\author{Romain Vasseur}
\affiliation{Department of Physics, University of Massachusetts, Amherst, MA 01003, USA}
\author{Andrew C. Potter}
\affiliation{Department of Physics and Astronomy, and Quantum Matter Institute,
University of British Columbia, Vancouver, BC, Canada V6T 1Z1}
\begin{abstract}
Monitored quantum circuits (MRCs) exhibit a measurement-induced phase transition between area-law and volume-law entanglement scaling. MRCs with a conserved charge additionally exhibit two distinct volume-law entangled phases that cannot be characterized by equilibrium notions of symmetry-breaking or topological order, but rather by the non-equilibrium dynamics and steady-state distribution of charge fluctuations. These include a charge-fuzzy phase in which charge information is rapidly scrambled leading to slowly decaying spatial fluctuations of charge in the steady state, and a charge-sharp phase in which measurements collapse quantum fluctuations of charge without destroying the volume-law entanglement of neutral degrees of freedom. By taking a continuous-time, weak-measurement limit, we construct a controlled replica field theory description of these phases and their intervening charge-sharpening transition in one spatial dimension. We find that the charge fuzzy phase is a critical phase with continuously evolving critical exponents that terminates in a modified Kosterlitz-Thouless transition to the short-range correlated charge-sharp phase. We numerically corroborate these scaling predictions also hold for discrete-time projective-measurement circuit models using large-scale matrix-product state simulations, and discuss generalizations to higher dimensions.
% Quantum circuits that are subject to a global conservation law, as well as to random local mea- surements of the conserved charge density, undergo a “charge-sharpening” transition; this transition separates a “fuzzy” phase in which the uncertainty in the total charge decays slowly, from a “sharp” phase where it decays rapidly. We construct a replica field theory of this transition. In one spatial dimension, we find that the sharpening transition is in a Kosterlitz-Thouless universality class, and that the fuzzy phase is a critical phase with power-law correlations. We provide extensive numerical evidence for the latter claim. We discuss how our results generalize to higher dimensions.
\end{abstract}
\maketitle

\paragraph{Introduction} 
Understanding the role of environmental dissipation and decoherence in large quantum circuits is essential for understanding the fundamental capabilities and limitations of noisy quantum computation relevant for near-term devices. A key conjecture of quantum complexity theory is that the output of individual quantum circuits is quite generally exponentially-hard (in qubit number) to predict classically. 
Examples where a full microscopic descriptions are intractable abound in physical systems:
from tracking individual motion of thermodynamically-many colliding gas particles, to computing the complex level structure of chaotic nuclei. Here, the successful philosophy of statistical mechanics and random matrix theory has taught us that the statistical properties of \emph{ensembles} of such complex systems can be far simpler to describe than individual realizations, and in many cases exhibit beautifully universal, model-independent properties. 

This observation has motivated the search for an analogous statistical mechanics paradigm for ensembles of random quantum circuits~\cite{HaydenPreskill2007,2008Fastscramblers,Brown:2015aa,Hosur:2016aa,PhysRevLett.98.130502,Brandao:2016aa,PhysRevX.7.031016,Nahum2018}. 
For circuits with ideal (noiseless) gates, this has led to identification of universal features of entanglement growth~\cite{Nahum2018,PhysRevX.9.021033}, scrambling~\cite{Nahum2018,VonKeyserlingk2018,PhysRevX.8.031057,PhysRevX.8.031058}, and quantum chaos~\cite{PhysRevX.8.041019,PhysRevLett.121.264101,PhysRevLett.121.060601,Friedman2019,PhysRevB.100.064309,PhysRevLett.123.210601}. Further, exploring monitored random circuits (MRCs), where decoherence is modeled by random measurements of the system by its environment, has revealed the possibility of sharp measurement-induced phase transitions (MIPTs)~\cite{PhysRevB.98.205136,Skinner2019} in the quantum trajectories (i.e. states produced by an MRC for fixed measurement outcomes) between a scrambling regime dominated by unitary gate evolution, and purifying dynamics dominated by measurement induced collapse~\cite{PhysRevB.98.205136,Skinner2019,Li2019,PhysRevB.99.224307,Li2020,10.21468/SciPostPhys.7.2.024,Gullans2019,Szyniszewski2019,Choi2020,Bao2020,Jian2020,Gullans2020,Zabalo2020,PhysRevResearch.2.023288,Ippoliti2020,Lavasani2020,Sang2020,PhysRevResearch.2.013022,PhysRevB.102.064202,Nahum2020,Turkeshi2020,Fuji2020,Lunt, Lunt2020,fan2020self,2020arXiv200503052V,Li2020b,PhysRevB.103.224210,PhysRevLett.126.060501,2021arXiv210306356L,2020arXiv201204666J,PhysRevLett.126.170503,2021arXiv210106245T,2021arXiv210209164B,2021arXiv210413372B,2021arXiv210407688B,2021arXiv210703393Z,agrawal2021entanglement,2021arXiv210804274L,PhysRevLett.126.170602,2021arXiv210609635J,2020arXiv201203857L,2021arXiv210208381B,turkeshi2021entanglement}. These raise the intriguing prospect of using statistical mechanics tools~\cite{PhysRevB.100.134203,Zhou2019,Bao2020,Jian2020} to study quantum communication channel capacity~\cite{Gullans2019,PhysRevX.11.031066}, error-correction thresholds~\cite{Gullans2019,Choi2020,fan2020self,Li2020b,2021arXiv210804274L}, and computational complexity~\cite{2019arXiv190512053H,Napp2019}. Finite-size evidence for such entanglement MIPT was even recently observed experimentally in trapped-ion chains~\cite{noel2021}.

In physical systems, symmetries play a central role in determining universal properties of phase transitions and protecting stable and distinct phases of matter, and may be naturally expected to play an important role in measurement-induced phases and critical phenomena in MRCs. Indeed, there is accumulating numerical evidence that symmetries can give rise to multiple distinct phases and critical phenomena in the highly-entangled regime -- which would support only classical, incoherent orders in thermal equilibrium. To date, the study of such measurement-stabilized orders has largely resorted to numerical analyses on Clifford circuits~\cite{Ippoliti2020,Lavasani2020,Sang2020,2021arXiv210209164B}. 
 Here, building on a statistical mechanics mapping developed in Ref.~\cite{agrawal2021entanglement}, we construct a replica field-theory framework to analytically study phases and critical phenomena in the volume-entangled regime of MRCs. We focus on the application of this technique to studying charge-sharpening transitions in more generic $1d$ MRCs with a conserved $U(1)$ charge or spin, and show that this transition is captured by a modified Kosterlitz-Thouless transition, and validate this prediction against large-scale matrix-product state (MPS) numerics. 
 %In supplemental appendices, we show that simple generalizations of this framework allow a controlled analytic exploration of a wide array of symmetry-breaking and topological orders in MRCs.

\paragraph{Model of random circuits with symmetries}
We consider a model~\cite{PhysRevX.8.031057} consisting of a $1d$ lattice, with a charged qubit with charge-basis states $|q=\pm 1\>$ and a neutral $d$-level qudit on each site, that evolves under a ``brick wall'' circuit of nearest-neighbor gates that conserve the total charge of the qubit pair, but are otherwise Haar-random in each block of fixed total charge. We consider randomly-placed single-site projective measurements with probability $p$. These measurements occur in the charge basis of the qubits and an arbitrary basis of the qudits. %with probability $p$. 
As shown in Ref.~\cite{agrawal2021entanglement}, this model supports two types of phase transitions (separating three distinct dynamical phases): an area-to-volume law entanglement transition at $p=p_c$ (identical to that of asymmetric circuits), and a ``charge-sharpening" transition at $p=p_\#$ occurring within the volume-law entangled phases.
%~\footnote{See supplemental material for a general argument that charge sharpening precedes the entanglement transition: $p_\#<p_c$, consistent with the large-$d$ numerical results of Ref.~\onlinecite{agrawal2021entanglement}.} 
The charge-sharpening transition distinguishes a charge-fuzzy phase ($p<p_\#$) in which scrambling is able to ``hide'' quantum superpositions of total charge from the measurements for a time that diverges with the system size, %from measurements by the circuit scrambling dynamics, 
and a charge-sharp phase ($p>p_\#$) in which the measurements collapse quantum superpositions of different total charge at a finite-rate. Throughout both phases the neutral qudit degrees of freedom remain volume-law entangled.
\begin{figure}[t]
\centering
\includegraphics[width=0.95\columnwidth]{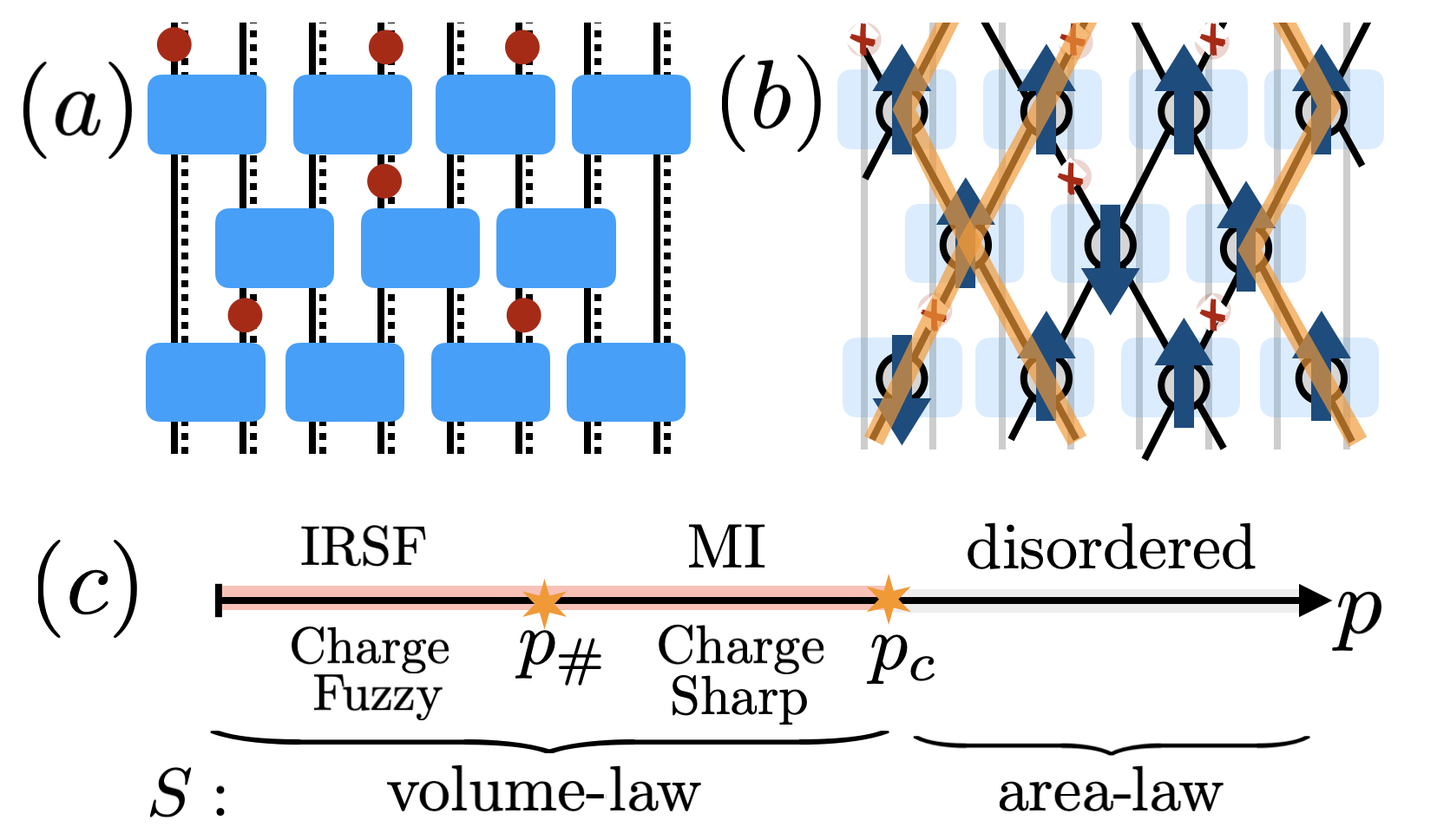}
\caption{\textbf{Model and phase diagram -- } (a) Monitored random circuit (MRC) with charged qubits (solid lines) and neutral large-d qudits (dashed lines) interacting with Haar-random gates (blue boxes) and randomly placed measurements (red dots). (b) Statistical mechanics (stat-mech) model for replicated MRC consists of replica permutation ``spins" (arrows) interacting with random-walking charge world-lines (orange lines). (c) The phase diagram, with phases labeled in the MRC (bottom) and stat-mech (top) language respectively. 
In addition to the entanglement transition at $p_c$, which corresponds to a (dis)ordering transition of the permutation ``spins", there is a charge-sharpening transition in the volume law phase, corresponding to an inter-replica superfluid (IRSF) to Mott-insulator (MI) transition in the statistical mechanics language. In $1{+}1d$, this charge sharpening transition has a modified Kosterlitz-Thouless universality class, and the IRSF is a critical/Goldstone phase with scaling exponents that vary continuously with $p$.
\label{fig:model} 
}
\end{figure}

As shown in Ref.~\onlinecite{agrawal2021entanglement}, the statistical properties of entanglement and charge correlators for this MRC ensemble can be captured, via a replica trick, by a classical statistical mechanics model defined on the graph of the quantum circuit (i.e. identifying gates with vertices and qubit world-lines between gates with links), and consisting of the following degrees of freedom:
 i) replica permutation ``spins" $s_i\in S_Q$ on each vertex $i$ where $Q$ is the number of replica copies, and 
 ii) charge degrees of freedom $q_{\ell, a} \in \{\pm 1\}$ on each link $\ell$ and replica $a=1\dots Q$.
As previously described in multiple works~\cite{PhysRevB.100.134203,Zhou2019,Bao2020,Jian2020} the entanglement transition at $p_c$ appears as an order/disorder transition of the permutation spins.
%, with entanglement of spatial region $A$ corresponding to the free-energy cost of a boundary domain wall between different ordered ``spin" configurations. 

Here, we focus on the charge-sharpening transition that occurs in the volume-law phase where the permutation spins remain ordered, and can be traced out to obtain a description purely in terms of the charge degrees of freedom. This can be done exactly in the limit of large qudit dimension $d$~\footnote{Since the permutation degrees of freedom are gapped for $p=p_\# < p_c$, we expect finite $d$ corrections to only renormalize the parameters of the effective field theory to be derived below.}. The resulting charge dynamics are then described by a classical stochastic process in which charge world-lines execute hardcore random walks in each replica. Measurements force the charges to coincide across replicas at the measured link, creating a space-time-disordered inter-replica interaction. 

These charge dynamics are described by a stochastic Markov process for the diagonal components of the (replicated) density matrix in the charge basis (with off-diagonal coherences strictly vanishing due to the qudit ``baths"). These form a $2^{LQ}$ component vector: $|\rho_Q\>$, which, if correctly normalized, satisfies $\<1|\rho_Q\>=\text{tr}\hat{\rho_Q}=1$ where $|1\>$ is the vector with all unit entries. The measurement- and gate- averaged evolution is described by a transfer matrix:
\begin{align}
|\rho_Q(t+1)\> = T_{\vec{m}(t+1/2)}T_{U,o}T_{\vec{m}(t)}T_{U,e}|\rho_Q(t)\>,
\label{eq:markov}
\end{align}
%where $T_{U,e/o} = \prod_{\<ij\>\in e/o} \prod_{a=1}^Q e^{J\(\vec{\sigma}_{a,i}\cdot\vec{\sigma}_{a,j}-\frac14\)}$  with $J\rightarrow\infty$ (i.e. each term is a projector onto the spin-triplet sector)
where $T_{U,e/o} = \prod_{\<ij\>\in e/o}\prod_{a=1}^Q \frac14\(\vec{\sigma}_{a,i}\cdot\vec{\sigma}_{a,j}+3\)$ project onto the spin-triplet sector for each bond and represent the evolution from random gates on even(e)/odd(o) bonds, and $\vec{\sigma}$ are Pauli matrices with $\sigma^z_{a,i}$ eigenvalue corresponding to the charge at link $i\in \{1\dots L\}$ in replica $a\in \{1\dots Q\}$. The measurement operators: $T_{\vec{m}(t)} = \prod_{i\in \mathcal{M}(t)}\prod_{a=1}^Q\delta_{\sigma^z_{a,i},m_i(t)}$ simply force the charges in all replicas to agree with the measurement outcomes $\vec{m}(t)$ on measured links $\mathcal{M}(t)$ at time-slice $t$. 

In the following, we will not work with explicitly normalized states, and use the replica trick to properly compute moments of local observables as:
%\begin{align}
%\mathbb{E}\[\<\hat{O}_1\>\<\hat{O}_2\>\] =  
%\mathbb{E}\[\frac{\<O_1|\rho_1\>\<O_2|\rho_1\>}{\<1|\rho_1\>^2} \<1|\rho_1\>\]
%\nonumber\\
% = \lim_{Q\rightarrow 1} \<O_1\otimes O_2\otimes 1\dots \otimes 1 \left\vert \mathbb{E}\[\rho_Q\]\right\rangle
% \label{eq:replica}
%\end{align}
%where $\mathbb{E}[\dots]$ denotes an average over trajectories, and $\<\dots\>$ denotes the quantum average within a trajectory, and we define $|O\>$ as the vector of diagonal (in the charge-basis) components of quantum operator $\hat{O}$: $\<n|O\> = \hat{O}_{nn}$. In the first line, the factors of $\<1|\rho\>$ in the denominator serve to explicitly normalize the state, and the extra factor of $\<1|\rho\>$ in the numerator weights each measurement outcome by its Born probability. 
%
\begin{align}
\mathbb{E}\[\<\hat{O}_1\>\<\hat{O}_2\>\] =  
\mathbb{E}\[\frac{\<1|O^{(d)}_1|\rho_1\>\<1|O^{(d)}_2|\rho_1\>}{\<1|\rho_1\>^2} \<1|\rho_1\>\]
\nonumber\\
 = \lim_{Q\rightarrow 1} \<1|O^{(d)}_1\otimes O^{(d)}_2\otimes \mathbbm{1}\dots \otimes \mathbbm{1} \left\vert \mathbb{E}\[\rho_Q\]\right\rangle,
 \label{eq:replica}
\end{align}
where $\mathbb{E}[\dots]$ denotes an average over trajectories, and $\<\dots\>$ denotes the quantum average within a trajectory, an we define the diagonal part of a quantum operator $\hat{O}$ as $O^{(d)} = \sum_m \<m|\hat{O}|m\> |m\>\<m|$ where $|n\>$ is a basis state with definite $\hat{\sigma}^z_{a,i}=m_{a,i}$. In the first line, the factors of $\<1|\rho\>$ in the denominator serve to explicitly normalize the state, and the extra factor of $\<1|\rho\>$ in the numerator weights each measurement outcome by its Born probability. 

In Ref.~\cite{agrawal2021entanglement}, this transfer-matrix model was analyzed explicitly using exact diagonalization (ED) methods. Here, we benchmark the field-theory predictions representing $|\rho\>$ as a matrix product state (MPS) using time-evolving block decimation (TEBD) analysis of Eq.~\ref{eq:markov}~\cite{PhysRevLett.91.147902,PhysRevLett.93.040502,SCHOLLWOCK201196}. We emphasize that in this statistical mechanics-description, the volume law phases of the physical qubits correspond to area-law (with log-violation for $p<p_\#$) phases of the statistical mechanics ``spins", $\vec{\sigma}$, enabling us to obtain results on much larger systems than ED (up to $\approx 60$ sites~\footnote{While much larger critical systems with 100's or 1000's of sites can be simulated by MPS methods in clean $1d$ systems, here, the disordered nature of the model adds significant sampling complexity and limits the achievable system size.}).

\begin{figure*}[t]
\centering
\includegraphics[width=2.0\columnwidth]{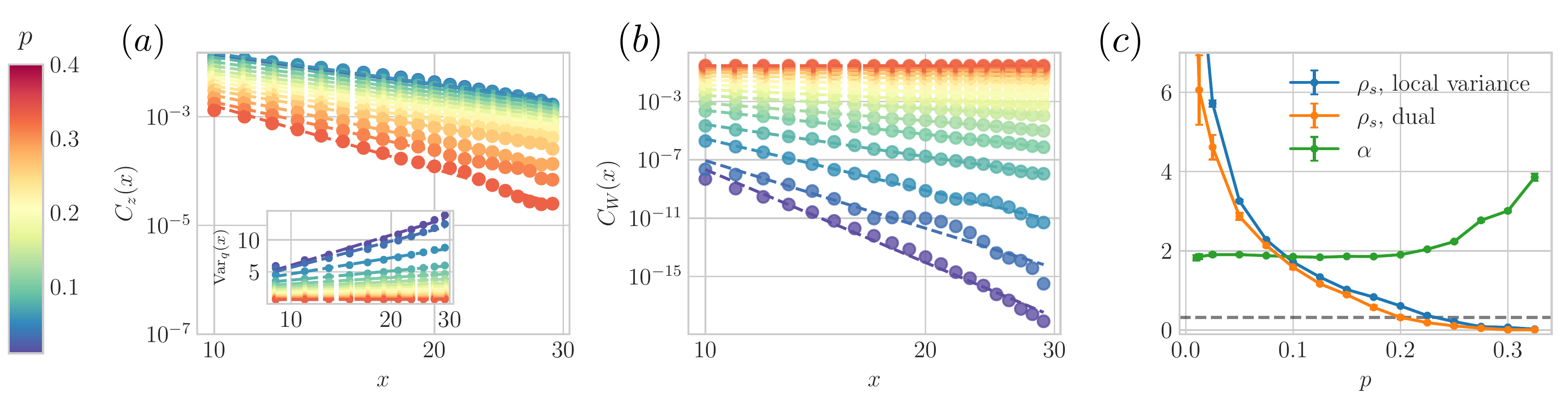}
%\vspace{1.5in}
\caption{\textbf{TEBD data -- } (a) Charge fluctuations $C_{z}(x)=\E\[\<\sigma^z_x\sigma^z_0\>-\<\sigma^z_x\>\<\sigma^z_0\>\]$, scaling as $C_{z}(x)\sim x^{-\alpha}$ in the fuzzy phase. Inset: charge variance of an interval of size $x$ , $\text{Var}_q(x) = \sum_{0<i,j<x}\E\[\<\sigma^z_i\sigma^z_j\>_c\]$, predicted to scale as $\sim \frac{8\rho_s}{\pi} \log x$.  (b) Dual string disorder parameter $C_W(x) =\E\[\<W_{[0,x]}\>^2 \]$ with $W_{[0,x]}=\prod_{0<i<x} \sigma^z_i$, showing power-law decay $C_W(x) \sim x^{-2 \pi \rho_s} $ in the charge-fuzzy phase.  (c) Continuously evolving superfluid density $\rho_s$ as a function of $p$, extracted from the local charge variance (blue) and the dual correlator $C_W(x)$ (orange). The dashed horizontal line indicates the critical threshold $\(\rho_s\)_\# = \pi^{-1}$. For $p<p_\# \sim 0.2$, the charge correlator $C_{z}(x)$ decays with an exponent $\alpha=2$ (green). 
\label{fig:TEBD} 
}
\end{figure*}

\paragraph{Effective field theory}
To gain an analytic handle on the charge dynamics, define a continuous time version of the stroboscopic/circuit evolution of Eq.~\ref{eq:markov}, by replacing
% $J\rightarrow J \delta t$ in the 
the spin-triplet projectors in the $T_U$ terms with a ferromagnetic interaction $\frac14\(\vec{\sigma}_i\cdot\vec{\sigma}_j+3\)\rightarrow e^{J\vec{\sigma}_i\cdot\vec{\sigma}_jdt}$, and replacing sharp projective measurements in $T_{\vec{m}}$ by Gaussian-softened ``weak" measurements: $\delta_{\sigma^z_{a,i},m_{i}(t)}\rightarrow \exp\[-\frac\gamma2\sum_a\( \sigma^z_{a,i}-m_i(t)\)^2\]$, where $J$ and $\gamma$ are now treated as adjustable parameters that respectively control the strength of unitary gate evolution and measurements respectively. 
We note that a similar strategy was used in~\cite{2021arXiv210209164B} to study $\Z_2$-symmetric circuits with $Q=2$ replicas. Here, we will use the large-$d$ qudits to take the proper replica limit and recover exact scaling results.

Averaging over measurement outcomes, the transfer matrix for time $t$ then takes the form of imaginary time with respect to a lattice Hamiltonian: $T(t) = e^{-tH}$
with:
\begin{align}
H = -J\sum_{\<i,j\>;a}\vec{\sigma}_{a,i}\cdot\vec{\sigma}_{a,j} + \frac\gamma2\sum_{i;a,b}\sigma^z_{a,i}\Pi_{ab}\sigma^z_{b,i},
\label{eq:H}
\end{align}
where $\Pi_{ab}=\(\delta_{ab}-\frac{1}{Q}\)$ is a projector onto replica-asymmetric modes.

Without measurements ($\gamma=0$), the random circuit dynamics simply takes the form of imaginary time evolution with $SU(2)$ invariant Heisenberg ferromagnet dynamics. The long-time steady states (ground-states of $H$) are simply equal weight superpositions over all charge configurations with each fixed total charge. The elementary excitations of $H_{\gamma=0}$ (corresponding to decaying perturbations to the steady-state) are magnon excitations with dispersion (wave-vector dependent decay rate) $\e_k\sim Jk^2$. These simply reflect the diffusive relaxation dynamics of conserved charges.
%, and the $SU(2)$ invariance of the magnon modes are essential for reproducing this feature. 
%
Measurements penalize differences in $\sigma^z$ between different replicas. After averaging over the space-time quenched disorder due to measurement locations and outcomes, this introduces inter-replica interactions and produces an easy-plane anisotropy for the inter-replica modes.  

We next construct an effective field theory by writing $T(t)$ as a spin-coherent state path integral in terms of polar angles $\theta_{i,a}$ and azimuthal angles $\phi_{i,a}$ for each spin~\cite{suppmat}.  For specificity, we work near zero charge density $\theta =\pi/2 +\delta \theta$.  Integrating out all $(Q-1)$ components of the out-of-plane fluctuations in the inter-replica modes, $\Pi\theta$, which are massive for any $\gamma>0$, and performing a fluctuation and gradient expansion gives an effective action 
%$\<1|T(t)|\rho_Q\> = \mathcal{Z}_Q = \int D\bar\theta D\bar\phi D\Pi\delta\theta D\Pi\phi e^{-\int_0^t dt \int dx \mathcal{L}_\text{eff}}$ 
$\<1|T(t)|\rho_Q\> = \mathcal{Z}_Q = \int D[\theta,\phi] e^{-\int_0^t dt \int dx \mathcal{L}_\text{eff}}$ 
with:
\begin{align}
\mathcal{L}_\text{eff} =& \frac{i}{2}\delta\bar\theta\d_t\bar\phi+\frac{\bar\rho}{2}\[\(\d_x\delta \bar\theta\)^2+\(\d_x\bar\phi\)^2\]
+\frac{\rho_s}{2}\(\d_\mu\Pi\phi\)^2,
\label{eq:Seff}
\end{align}
where $\mu\in \{t,x\}$, and repeated indices are implicitly summed, $\bar\rho \sim J$, $\rho_s \sim \sqrt{J/\gamma}$, and we have defined the replica average modes: $\bar\phi,\bar\theta \equiv \frac1Q\sum_{a=1}^Q \phi_a,\theta_a$~\footnote{More generally, $\rho_s$ depends on the space-time local charge density.  But this coupling is RG-irrelevant in $1{+}1d$ so we drop it here, although it can become relevant in higher $d$, as we discuss later.}. To compute correlators as in Eq.~\ref{eq:replica}, this action should be supplemented by boundary conditions corresponding to the final state $\<1|$ which is an equal weight superposition of all charge states, corresponding to a product state of spins point along the $\hat{x}$ direction: $(\theta,\phi)=(\frac\pi2,0)$ at the final time, $t$.
%Since only charge-diagonal observables contribute after trajectory averaging, the most general form of $|O_i\>$ is obtained by rotating $|1\>$ by some linear superposition of $\sigma^z_{a,i}$ operators which can be written as $\<O_a|=\<1|e^{i\sum_{i}\varphi_{a,i}\sigma^z_{a,i}}$, which corresponds to boundary conditions $\phi_{a}(x,t)=\varphi_{a}(x)$. As a particular example, if 
In particular, steady-state ($t\sim L\rightarrow\infty$) correlators are generated from the partition function on the half- plane $(t,x)\in\(-\infty,0\]\times\mathbb{R}$ with boundary conditions $\phi_a(t=0,x)=0$ and appropriate (charge-diagonal) operators inserted.
As a consequence, steady-state properties of MIPTs will correspond to \emph{boundary-critical properties} of the statistical mechanics problem.

The ``replica-average" modes $\bar\phi$ determine simple linear observable averages, $\bar{O}=\E[\<\hat{O}\>]$. These are unaffected by measurements and have a simple FM spin-wave action and converge to $\tr\hat{O}$ at late times independent of $\gamma$. This accords with the well-known fact that MIPTs are only visible in higher-moments and non-linear functions of state. 

The inter-replica fluctuations, $\Pi\phi$, control disconnected moments of correlation functions such as $\E\[\<O(x)O(0)\>-\<O(x)\>\<O(0)\>\]$. When singular vortex configurations in the phase-fields are irrelevant ($0<\gamma<\gamma_{\#}$) the inter-replica modes follow a superfluid action with $(Q-1)$ decoupled relativistic Goldstone-mode excitations which indicate that charge fluctuations with wave-vector $k$ decay at rate $\sim |k|$  (dynamical exponent $z=1$). This inter-replica-superfluid (IRSF) phase represents the charge-fuzzy phase ($0<p\leq p_\#$ in the circuit model).
Since the effective ``superfluid stiffness'' $\rho_s$ decreases monotonically with increasing measurement strength, $\gamma$, it is natural to expect that the charge-sharpening transition in $1{+}1d$ is a Kosterlitz-Thouless (KT)-type transition where vortex-proliferation destroys the IRSF QLRO for $p>p_\#$, i.e. $\gamma>\gamma_\#$, resulting in a ``Mott insulating'' phase. This picture will turn out to be qualitatively correct, albeit with important quantitative changes to the usual KT transition due to the replica structure.

%Equivalently, these results imply that the leading wave-vector, $k$, dependence of the steady-state charge-structure factor (Fourier transform of Eq.~\ref{eq:chargecorrelators}) changes from non-analytic, $C(k)\sim |k|$, in the fuzzy phase and at the sharpening transition, to analytic $C(k)\sim k^2$ in the charge-sharp phase.
%\red{Should we give David's hand-wavy argument for this here or in an appendix?}

\paragraph{Charge-sharpening in $1{+}1d$}
To obtain a controlled theory of the transition, we introduce vortex defects into Eq.~\ref{eq:Seff} by standard duality methods~\cite{suppmat} to obtain modified ``sine-Gordon'' model:
\begin{align}
\mathcal{L}_\text{dual} =& \frac{1}{8\pi^2 \bar\rho}\[\(\d_t \bar\vartheta\)^2+D^2 \(\d_x^2\bar\vartheta\)^2\]+ 
\nonumber\\
&+ \frac{1}{8\pi^2\rho_s}(\Pi\d_\mu\vartheta)^2
-\lambda\sum_{a\neq b}\cos(\vartheta_a-\vartheta_b),
\label{eq:Ldual}
\end{align}
where $D \sim J$, $e^{-i\vartheta_a}$ inserts a (spacetime/instanton) vortex, and $\vartheta$ are related to the original fields by $\rho_s\d_\mu \phi_a \leftrightarrow \frac{\epsilon^{\mu\nu}}{2\pi}\d_\nu\vartheta_a$, $\lambda\approx e^{-\sqrt{J/\gamma}}$ is the vortex fugacity, and we have kept only the most relevant vortex terms.
Note that the minimal topological defects that can appear are actually a bound states of a vortex and anti-vortex in different replicas. Formally, this is because individual vortices, which contribute vorticity to $\bar\phi$, are linearly confined by the diffusive replica-average mode. Intuitively, this simply reflects the absence of quantum fluctuations in the Heisenberg ferromagnet ground-state that describes replica-averages in the steady-state. An immediate consequence of this vortex-``doubling'' is that it \emph{halves} the critical superfluid stiffness compared to the ordinary KT transition:
$\(\rho_s\)_\# = \pi^{-1} = \frac12 \(\rho_s\)_\text{KT}$.
We further note, that in an ordinary superfluid, vortex condensation requires commensuration between particle density and the lattice, otherwise vortex instantons acquire non-trivial Berry phases and are suppressed. Here, the density conjugate to the composite vortex operators are inter-replica density fluctuations, which has vanishing average independent of the physical (replica-average) charge density. Consequently, in $1{+}1d$, there is a single universality class for charge-sharpening, in contrast to the ordinary superfluid-Mott transition which arises only at integer densities and exhibits different scalings in the presence or absence of particle-hole symmetry.

\paragraph{Observables and numerics}
In $1{+}1d$, the fuzzy-phase/IRSF exhibits only quasi-long-range order (QLRO), with algebraic decay of charge ($\sigma^z \approx \frac{1}{\pi}\d_x\vartheta$) correlators, $C_z(x) = \E\[\<\sigma^z(x)\sigma^z(0)\>-\<\sigma^z(x)\>\<\sigma^z(0)\>\]$, which are negative at large distance in the steady state.  This changes to short-range correlations in the gapped phase: 
\begin{align}
C_{z}(x) \sim -
\begin{cases}
\rho_s(a/x)^{-2} & p\leq p_\# \\
e^{-x/\xi} & p> p_\#
\end{cases}
+\dots
\label{eq:chargecorrelators}
\end{align}
where $a$ is a non-universal UV cutoff (lattice spacing), $\xi$ is a finite correlation length and $(\dots)$ denote asymptotically subleading terms. This behavior is consistent in TEBD results for the discrete-time model Eq.~\ref{eq:markov} showing an algebraic decay of $\E[\<\sigma^z(x)\sigma^z(0)\>_c]$ with power-law fit  that is constant over an extended range $0<p<p_\#\approx 0.2$ (Fig.~\ref{fig:TEBD}a,c).

A hallmark of KT-physics is that certain correlators exhibit continuously evolving critical exponents in the QLRO Goldstone-phase. Constraining ourselves to charge-diagonal quantities that can be physically probed in the original qubit language, a convenient observable that displays this behavior are the string operators:
$W_{[0,x]} = e^{-i\pi \sum_{i\in(0,x)}\sigma^z_i/2} \approx e^{-i\frac{1}{2}\int_0^x d\vartheta } = e^{-i\vartheta(x)/2}e^{i\vartheta(0)/2}$, which inserts a $\pi$-phase twist in the $\phi$-fields in the interval $[0,x]$, and can be thought of as a dual (boundary) order parameter for vortex condensation: 
\begin{align}
C_W(x) = \E\[\<W_{[0,x]}\>^2\] \approx
\begin{cases}
|x|^{-2\pi \rho_s} & p\leq p_\# \\
 \text{constant} & p>p_\#
 \end{cases}.
\end{align}
The scaling dimension of $W$ decreases monotonically with measurement rate $\gamma$ (as $\sim \gamma^{-1/2}$ for small $\gamma$), and jumps (for $L\rightarrow \infty$) discontinously to $0$ in the charge-sharp phase ($\gamma>\gamma_\#$), achieving a minimum non-zero value of $\Delta_W = 1$ at the sharpening transition ($\gamma=\gamma_\#$). 

%, and strongly suggest that the charge-sharpening transition at $p_\#\approx 0.15??$ occurs well-before the value, $\approx 0.3$ estimated in Ref.~\cite{agrawal2021entanglement} -- a well-known pitfall of applying single-parameter scaling analysis to two-parameter-like KT transitions which have formally divergent correlation length exponent, $\nu=\infty$.

%This prediction is supported by TEBD simulations which show a continuously evolving exponential decay for the lattice-version of $W\rightarrow \prod_{0<i<x}(2\sigma^z)$, with fitted exponent decreasing monotonically to $\sim -2$ at a value of $p \approx 0.15??$ which we identify with $p_\#$. This value of $p_\#$ is reasonably consistent with the end $\alpha=2$ plateau the decay exponent of $\sigma^z$-correlators. We note that $p_\#$ is significantly reduced compared to the value $\approx 0.3$ previously identified in~\cite{} by scaling collapses of ED data with a finite correlation exponent $\nu$~\footnote{Such overestimates of the critical point are very common when applying standard one-parameter scaling analysis to KT transitions.}

The predicted power-law decay of charge- and string- correlators are in excellent agreement with TEBD data (Fig.~\ref{fig:TEBD}) for $0<p\leq p_\#\approx 0.2$. We note that, as is typical for two-parameter scaling KT-transitions, incorrectly applying a single-parameter scaling analysis with finite correlation-length exponent, $\nu$ as in Ref.~\cite{agrawal2021entanglement} dramatically overestimates the critical measurement strength, and misses the key physics of continuously evolving scaling exponents in the charge-fuzzy phase.

\paragraph{Discussion}
The agreement with unbiased numerics gives strong support to the hypothesis that the continuum field theory description captures the universality class of charge sharpening in the discrete-time, strong-measurement model. Importantly, the analytic replica field theory approach can be applied in more complicated scenarios where numerics becomes challenging.  In the supplemental material~\cite{suppmat}, we give a general argument that the charge-sharpening transition occurs separately from, and precedes the entanglement transition in any spatial dimension $D$, and show that for $U(1)$-symmetry is again described by $(Q-1)$ coupled XY-models for the inter-replica modes. A new feature of $D\geq 2$ $U(1)$-MRCs is that, since circuit-evolution time plays the role of temperature in the stat-mech model, for $p<p_\#$, the system exhibits a finite-time transition at critical time $t_c$, where the long-range correlations appear, although the charge remains ``fuzzy'' until a time $\sim L^D$. For $D\geq 2$ this finite-time transition has the same critical properties as a superfluid transition at nonzero temperature in D dimensions. This critical time diverges as $t_c\sim |p-p_\#|^{-z/\nu}$ with $z=1$ and $\nu=\nu_{D+1}$. Evidence and arguments for related finite-time complexity~\cite{Napp2019} and entanglement~\cite{2021arXiv211006963B} transitions in MRCs have previously been discussed absent symmetries, however, in the present case, the field theory approach gives a controlled calculation of critical properties in generic Haar-random circuit models. A second twist that arises in $D\geq 2$ is the emergence of multiple classes of boundary-criticality governing the steady-state behavior of MRCs: i) ordinary (bulk orders before boundary), ii) extra-ordinary (boundary orders before bulk), and iii) special (boundary transition between ordinary and extraordinary), depending on the relative bulk and boundary coupling strengths. Here, the extra-ordinary to ordinary boundary transition can be tuned by increasing the number of measurements in the final few circuit layers compared to the bulk. The properties of the extraordinary transition for the $2{+}1d$ XY model were worked out only very recently~\cite{2020arXiv200905119M}, and exhibits an unusual log-divergence of $\rho_s$ upon approaching the bulk phase transition with the QLRO surface. A third new feature that arises in $2{+1}d$ and $3{+1}d$ when we move away for mean charge density zero is a relevant coupling between the charge diffusion and the inter-replica dynamics via the dependence of $\rho_s$ on the local charge density, which moves the sharpening transition to a new universality class with quenched disorder and, presumably, $z>1$.

Finally, it is straightforward to generalize the field theory approach to study charge-sharpening transitions with discrete symmetries. Presumably the main difference from continuous symmetry will be that there are no gapless Goldstone modes in the charge-fuzzy phase, so that correlators are either long-range or exponential decaying.  By way of standard duality transformations~\cite{agrawal2021entanglement}, this implies the existence of sharp symmetry-breaking and topological-ordering/(de)confinement transitions \emph{within the volume law entangled regime} whereas only classical orders would be allowed in thermal equilibrium. The prospect of a quantum-ordered phases in volume-law entangled trajectories was discussed in various Clifford models~\cite{2021arXiv210209164B}. Our field-theory technique not only places this discussion on firmer ground for generic (Haar-random) MRC classes, but also enables controlled calculation of symmetry-breaking and topological MIPTs. We expect that this statistical mechanics field theory approach might also be useful for studying more complicated transitions arising from the competition between non-commuting measurements. It would also be highly desirable to obtain a controlled field-theory description of the entanglement transition absent symmetries -- though at present this appears a daunting challenge.

{\it Acknowledgements -- } We thank A. Zabalo, K. Chen, M.P.A. Fisher, M. Gullans, J. Pixley and J. Wilson for useful discussions and/or earlier collaborations on charge sharpening transitions. The tensor network simulations  are based on the quimb library~\cite{gray2018quimb}. We acknowledge support from NSF DMR-1653271 (S.G.), NSF DMR-1653007 (A.C.P.), NSF QLCI grant OMA-2120757 (D.A.H.), the Air Force Office of Scientific Research under Grant No. FA9550-21-1-0123 (F.B. and R.V.), and the Alfred P. Sloan Foundation through Sloan Research Fellowships (A.C.P. and R.V.).

\bibliography{references}
\bibliographystyle{apsrev4-1}

\onecolumngrid
\appendix
\section{Derivation of Effective Field Theory}
\subsection{Statistical Mechanics Model (review)}
We briefly review the mapping derived in Ref.~\onlinecite{agrawal2021entanglement} from the charge-qubit $\times$ neutral-qudit circuit model to a statistical mechanics model, focusing on the large-d limit and volume-law entangled regime in which replica permutation spins form a ferromagnet and can be effectively dropped from the description without effecting universal results. We will then take the continuous time limit of this model to obtain the field theory analyzed in the main text.

We start by focusing on a single Haar-random two site gate with unitary that is block diagonal in the charge basis: 
\begin{align}
U = \bigoplus_{q\in \{-1,0,+1\}} U^{(q)}P_q = 
\begin{pmatrix}
U^{(-1)}_{d^2\times d^2} & 0 & 0\\
0 & U^{(0)}_{2d^2\times 2d^2} & 0 \\
0&0&U^{(+1)}_{d^2\times d^2}
 \end{pmatrix},
\end{align}
where $U^{(q)}_{N\times N}$ denote $N\times N$ Haar-random matrix blocks with total charge $q$, and $P_q$ projects onto the sector with total charge $q$.

As derived in~\onlinecite{agrawal2021entanglement}, taking $Q$ replicas, and averaging over Haar-random gates gives (for $d\rightarrow \infty$):
\begin{align}
\E\[U^Q\otimes U^{*Q}\]_{n_1,n_2}^{m_1,m_2} \underset{(d\rightarrow\infty)}{=} \sum_{\sigma \in S_Q}\sum_{q_a\in \{\pm 1,0\}}\bigotimes_{a=1}^Q\frac{1}{D_{q_\alpha}}\delta_{q_a,q'_{\sigma(a)}}\delta_{n_1,\sigma(m_1)}\delta_{n_2,\sigma(m_2)},
\label{eq:haaravg}
\end{align}
where $D_q = \begin{cases} d^2 & |q|=1 \\ 2d^2 & q=0 \end{cases}$ is the  dimension of the block with total charge $q$, $q_a$ is the total charge of the qubits in replica $a$ entering the gate, $q'_a$ is the total charge of the qubits leaving the gate, $S_Q$ is the symmetric group of permutations on $Q$ elements, and $n_{i},m_{i}$ are $Q$-component vectors specifying the charge state on the $i=1,2$ in replica $a$.

When computing charge observables (as opposed to entanglement observables), we may trace out the replica spin, $\sigma$ degrees of freedom. In the large-d limit and volume-law entangled phase, these form an ordered state, which we can take without loss of generality to be along the identity permutation $\sigma(a)=a$. 

In this limit, the matrix elements in Eq.~\ref{eq:haaravg} have a simple interpretation: given an incoming set of link charges flowing into a gate, any output charge configuration consistent with separate charge-conservation in each replica is equally likely. The partition function for a unitary quantum circuit made out of such Haar random gates takes the form of a disordered 6-vertex model~\cite{agrawal2021entanglement}, in which each gate is replaced by a vertex, for which each gate maximally randomizes the charges of its input legs. Projective measurements acting on a link simply force the charges to agree across all replicas. Their random position and measurement outcomes act as a source of space-time disorder.

An important feature of this large-$d$ limit, is that the Haar average locks the indices of $U$ and $U^*$ together, so that the Haar-averaged channel only propagates components of the density matrix that are \emph{diagonal} in the charge basis, and a single Haar-averaged gate layer will kill off-diagonal coherences. For this reason, we can reduce the replicated density matrix for the quantum system to a vector of it's diagonal components, $\hat{\rho} \rightarrow |\rho\>$ with $\<n|\rho\> = \<n|\hat{\rho}|n\>$.

  Combining these elements, one obtains the transfer-matrix description for the discrete time evolution of $|\rho_t\>$ written in the main text. 
%The partition function for this replicated transfer matrix takes the form of a disordered 6-vertex model~\cite{}, where the charge configurations exiting a vertex are an equal-weight superposition of all configurations consistent with the total incoming charge. The randomly space-time locations and measurement outcomes introduced space-time random disorder that couple together charges on different replicas. 
In the next section, we describe a prescription to construct a continuous-time relaxation of this discrete stochastic evolution, which facilitates a continuum field theory approach.

\subsection{Continuum Time Limit}

\paragraph{Unitary evolution} 
In the 6-vertex model, there is an equal probability that a charge word-line entering a vertex exits in either direction. A complementary, equivalent interpretation of the continuum time limit taken is that the word-line of a charge entering a vertex continues along the same leg it enters with probability $1-\frac{J}{2}dt$, and change directions with probability $\frac{J}{2}dt$ where $J$ is a coupling constant. This is implemented by replacing each vertex weight in the 6-vertex model by two-site transfer operator:
\begin{align}
T_{U,\text{2-site}} &= \frac{1+\sigma^z\otimes \sigma^z}{2}+\(\frac{1-\sigma^z\otimes \sigma^z}{2}\)\[(1-\frac{J}{2}dt)+\frac{J}{2}dt(\sigma^+\otimes\sigma^-+h.c.)\]= e^{J \vec{\sigma}_i\cdot\vec{\sigma}_j dt},
\end{align}
which takes the simple ferromagnetic Heisenberg form quoted in the main text.

\paragraph{Measurements}
In lieu of projective measurements, we define continuous (``weak") measurements through a term:
\begin{align}
H_M(\tau) = \frac{\gamma}{2} \sum_{i,a} (\sigma^z_{i,a} - m_i(\tau))^2,
\end{align}
which softly projects $\sigma^z$ onto measurement outcome $m_i$, which recovers the projective limit for $\gamma\rightarrow \infty$, these become projective measurements. We also generalize the measurement outcomes to be a continuous number, rather than $m_i\in \pm 1$. This softening of the measurement quantization is typical of standard theory of weak-measurements, and we do not expect these kind of details to change the universality class since the Gaussian weights tend to pin $m_i$ to allowed values of $\sigma^z$.

This formulation of weak measurements omits the spacetime randomness in the position of the measurements in the original circuit model, replacing them with a uniform weak measurement field. Spacetime randomness in $\gamma$ can also be introduced, but this turns out to be RG irrelevant in 1+1d, so we choose not to include it. Physically, we expect that this irrelevance may hold in all dimensions, since, in a given trajectory, space-time randomness should be self-generated by the (weak) measurement outcomes.

We note that, with the proper normalization enforced by the replica limit described above, this approach is very similar to the standard formulation of weak-measurements, which was recently used to construct field theories of MIPTs in non-interacting fermion systems~\cite{2021arXiv210208381B}. There measurements weight trajectories by $e^{-\frac\gamma 2\(\sigma^z-\<\sigma^z\>\)}$. Compared to our approach, this corresponds to replacing $m_i$ self-consistently with its average, which for is us approximately accomplished through the Born weights in the replica limit $Q\rightarrow 1$. This replica trick replaces the complicated task of performing non-linear evolution using state-dependent jump operators, which we will see enables a controlled analytic analysis for the generic (Haar-random) MRC dynamics considered in our paper.

It is convenient to shift the $m$-field to absorb the mean of $\sigma^z_i$ over replicas:
$H_M(\tau) \rightarrow \frac{\gamma}{2} \sum_{i,a} (\delta \sigma^z_{i,a} - m_i(\tau))^2
$, where $\delta \sigma^z_a = \sigma^z_a-\frac1Q\sum_b \sigma^z_b$. Noting that $\sum_a \delta \sigma^z_a =0$. After integrating out $m$, this gives effective Hamiltonian term:
\begin{align}
H_M &= \frac{\gamma}{2} \sum_{i,a}\(\sigma^z_{i,a} -\frac1Q\sum_b \sigma^z_{i,b}\)^2
= \frac\gamma2 \sum_{i,a,b}\sigma^z_{i,a} \(\delta_{a,b}-\frac{1}{Q}\)\sigma^z_{i,b}.
\end{align}

Combining the unitary and measurement parts yields the Hamiltonian of the main text. Next, we introduce a spin-coherent state path integral for transfer operator $T(t)$ which takes the form of imaginary time evolution with respect to $H$ for time $t$.

\paragraph{Spin coherent state path integral} Following standard prescription for constructing a path integral using spin-coherent states described by unit vector $\vec{\sigma}_i\rightarrow \hat\Omega_i$ written in terms of polar angle $\theta$ and azimuthal angle $\phi$, gives effective action:
\begin{align}
L =& -\frac i2\sum_{a,i} \cos\theta_{a,i} \d_\tau\phi_{a,i} 
-J\sum_{\<i,j\>,a}\[\cos\theta_{a,i}\cos\theta_{a,j}+\sin\theta_{a,i}\sin\theta_{a,j}\cos\(\phi_{a,i}-\phi_{a,j}\)\]
\nonumber\\
&
+\frac{\gamma }{2}\sum_{i;a,b}\cos\theta_{a,i}\Pi_{ab}\cos\theta_{b,i}.
\label{eq:Llattice}
\end{align}
where we have defined the projector onto inter-replica modes, $\Pi = \delta_{a,b}-Q^{-1}$.

Next, consider average particle density $n_0 = \cos\theta_0$. Then, expanding in small fluctuations around this mean value $\delta\theta=\theta-\theta_0$, and in the azimuthal angle $\phi_i-\phi_j \approx \hat{e}_{ij}\cdot\nabla \phi$ where $\hat{e}_{ij}$ is a unit vector along link $ij$, and converting lattice sums into continuous integrals over space gives Lagrangian density:
\begin{align}
	\mathcal{L} \approx& i\frac{\sin\theta_0}{2}\sum_{a} \delta\theta_{a,i} \d_\tau\phi_{a} 
		+\frac{J}{2}\sum_{i,a}\[\(\d_x\delta\theta_{a}\)^2+\sin^2\theta_0\(\d_x\phi_{a,i}\)^2\]
		+\frac{\gamma\sin^2\theta_0}{2}\sum \delta\theta_{a}\Pi_{a,b}\delta\theta_{b}
		\nonumber\\&+
		\frac{J}{2}\cos\theta_0\delta \theta_a\Pi_{ab}\(\nabla\phi_b\)^2 ,
\end{align}
where in the second term, we used $\cos\theta\cos\theta'+\sin\theta\sin\theta'=\cos(\theta-\theta')\approx 1-\frac12(\theta-\theta')^2$, dropped constant term(s), and ignored cross terms like $(\nabla\delta\theta)^2(\nabla\phi)^2$. Dropping higher-derivative terms is valid when vortex configurations are irrelevant. The effect of vortices will be considered below through a continuum duality, but could also be directly derived from Eq.~\ref{eq:Llattice} using standard lattice-duality mappings.

We next integrate out the $\Pi \delta \theta$ fluctuations, and write: $\phi_a = \Pi_{ab}\phi_b+\bar\phi$ to obtain:
\begin{align}
\mathcal{L} =& \frac{1}{8\gamma} \d_t\phi_a\Pi_{ab}\d_t\phi_b + 
\frac{J\sin^2\theta_0}{2}\nabla\phi_b\Pi_{ab}\nabla\phi_b 
\nonumber\\&
+\frac{i}{2}\sin\theta_0 \delta\bar\theta\d_t\bar\phi 
+ \frac{J}{2}\[\(\nabla \delta\bar\theta\)^2+\sin^2\theta_0\(\nabla\bar\phi\)^2\]
\nonumber\\&
+\frac{J}{2}\cos\theta_0 \delta\bar\theta\sum_a\(\nabla\Pi_{ab}\phi_b\)^2.
\end{align}
The term in the third line represents a linear coupling between the diffusive $\bar\theta$ fluctuations, and the energy-density of the $(Q-1)$ superfluid Goldstone modes. The slow, diffusion (dynamical exponent $z=2$) dynamics of the replica-average mode are effectively static on the time scale of the ballistic ($z=1$) inter-replica Goldstone mode fluctuations, which therefore see the last term as a static ``random-bond" disorder. For the Goldstone phase, such disorder is always irrelevant, since the Goldstone modes are stable. For a critical point, such random-bond disorder is relevant if $\nu<2/d$ (the Harris criterion~\cite{harris}). In $1+1d$, we will find a KT like transition with formally divergent $\nu=\infty$, hence this linear coupling of the diffusive and ballistic modes is irrelevant.  Near the transition in the Mott insulator, this randomness will induce rare-region Griffiths effects, making the Mott insulator gapless.  For $2+1d$, on the other hand, the natural generalization of the KT transition is a 3dXY transition, which has $\nu_\text{3dXY}\approx0.67<1$, and the diffusive and critical modes will not decouple. The same is true in $3+1d$, where the 4dXY transition has mean-field exponent $\nu=1/2<2/3$. This shows that, in dimensions higher than one, it is no longer possible to ignore the coupling of the replica-diffusive modes and the critical inter-replica fluctuations, and that this necessarily changes the universality class of the phase transition away from the simple XY universality class. An exception is for the particle-hole symmetric fillings, where $\theta_0=\pi/2$. In this case, the linear-coupling between diffusive and ballistic modes vanishes, and the next leading order is a quadratic coupling $\delta\bar\theta^2\(\nabla\phi_a\)^2$ is irrelevant if $\nu>1/d$, which is satisfied by the XY transition in all dimensions. However, it is possible that the inter-replica couplings could be relevant, so that a different universality class than XY occurs even for particle-hole symmetric fillings in dimensions $\geq 2$. We leave a detailed study of higher-dimensional transitions for future work, and return to $1+1d$.

To obtain the spacetime isotropic action, we rescale the spacetime coordinates $(x,t)\rightarrow (x',t') = (x/\sqrt{v},t\sqrt{v})$ where $v = 2|\sin\theta_0|\sqrt{\gamma J}$ is the velocity of the $\Pi\phi$ modes, and obtain:
\begin{align}
\rho_s = \frac{|\sin\theta_0|}{2}\sqrt{J/\gamma}.
\end{align}
As expected, we see that the superfluid stiffness, $\rho_s\sim \sqrt{J/\gamma}$ diverges in the limit of zero measurement strength (since there are no ``quantum fluctuations" in the spin at $\gamma\rightarrow 0$ in the Ferromagnet), and goes to zero at large $\gamma$, indicating that increasing the measurement strength should drive a disordering transition. 
Note that the critical measurement rate for charge sharpening scales like $\gamma_c\sim \sin^2\theta_0$
and vanishes as the completely full or empty states are approached ($\theta_0\rightarrow 0,\pi$ respectively). Intuitively, this reflects that nearly full or empty states have less dynamical fluctuations (e.g. have larger dead regions with no dynamics) and are easier to sharpen.

Henceforth, we consider particle-hole symmetric filling $\theta_0=\pi/2$, since, as we have argued above, this choice does not affect the universality class in $1+1d$.

%\red{revisit this to include replica labels} This action leaves out the action cost of the vortex-core (where the superfluid has to die off, i.e. the spins point along $\pm z$ in the core, which costs energy density $\sim \gamma$ over the core of spacetime volume $\sim $), which we can model by $\sum_{a,x} \e_c |n^v_a|^2$ where $\e_c \sim \sqrt{J/\gamma}$ {\it Note: this estimate is obtained by identifying the vortex core size $\xi_n$ for a vortex of vorticity $n$ by where the phase twist cost for the vortex at distance $\xi$ from the core: $\frac12\rho_s (\nabla\phi)^2 \sim \frac{n^2\rho_s}{\xi_n^2}$ is equal to the core action-density cost: $\frac12\sqrt{J\gamma}$, and then the action cost of the core is $\sim \sqrt{J\gamma}\xi_n^2\sim \sqrt{J/\gamma}n^2$}.

\subsection{Topological defects and duality}
To derive an effective action for vortices, decompose the phase fields into smooth (s) and vortex (v) parts: $\phi_a = \phi^s_a + \phi^v_a$. Next, we introduce a Hubbard-Stratonovich vector-field $j_{a}^\mu$ to decouple the $(\d_\mu\Pi\phi)^2$ terms, and $\(\d_x\bar\phi\)^2$ term. In addition, we redefine $\bar j_0 = \delta\bar\theta/2$  to obtain:
\begin{align}
\mathcal{L}[j,\phi] = ij^\mu_a\d_\mu\(\phi^s_a+\phi^v_a\) + \frac{1}{2 \bar\rho} \bar{j}_x^2 +2\bar\rho \(\d_x\bar{j}_0\)^2 + \frac{1}{2\rho_s} |\Pi_{ab} j^\mu_b|^2, 
\end{align}
Integrating out the smooth fluctuations in $\phi^s_a$ enforces current conservation: $\d_\mu j^\mu_a = 0$, which we can solve by introducing the dual vortex field:
\begin{align}
j^\mu_a = \frac{\epsilon^{\mu\nu}}{2\pi}\d_\nu \vartheta_a,
\end{align}
where $\epsilon^{\mu\nu}$ is the unit antisymmetric tensor. 

Next, integrating the $i\eps\d\vartheta \phi^v$ term by parts, and noting that $\epsilon^{\mu\nu}\d_\mu\(\d_\nu\phi_a\) = 2\pi \rho^v_a$ where $\rho^v_a$ is the space-time density of vortices in replica $a$, gives:
\begin{align}
\mathcal{L} = \frac{1}{8\pi^2 \bar\rho}\[  \(\d_t\bar\vartheta\)^2 +D^2 \(\d_x^2\bar\vartheta\)^2 \]+ \frac{1}{8\pi^2\rho_s} \(\Pi_{ab} \d_\mu \vartheta_b\)^2 + i\vartheta_a\rho^v_a,
\end{align}
with $D=2 \bar\rho$.

We next argue that vortices in the replica-average mode $\bar\phi$ should not be included in the low-energy effective description. First, the replica-average mode have full SU(2) rotation symmetry, which has trivial fundamental group, $\pi_1(SU(2))=1$. Hence replica-average vortices are not topologically stable and can decay into ferromagnetic magnons. Second, inter-replica vortex configurations are strongly confined, as can be seen by integrating out $\bar\vartheta$ in the above action in the presence of a replica-average vortex/anti-vortex pair separated by spacetime displacement $(t,x)$, which gives action cost $\sim \sqrt{|t|+x^2/D}$ that grows with distance much faster than the configurational entropy ($\sim \log \sqrt{x^2+t^2}$) of the vortices, and hence forbids these vortices from ever unbinding.

We then focus only on composite vortices with zero net vorticity across replicas. The minimal such configurations are pairs of $a$-vortex/$b$-antivortex with replica indices $a\neq b$. There are $Q(Q-1)$ species of these composite vortices. Each has the same fugacity due to replica permutation symmetry, and pairs of these composite vortices have mutual $2d$-Coulomb interactions with potential $\sim \log |x|$, when they share a flavor label.

The above derivation from the phase-only action leaves out the action cost of the vortex core, where the inter-replica superfluid must die off. Estimating the core size $\xi$ as the distance where the phase-gradient action-density cost becomes comparable to that of condensation, we estimate for inter-replica vortices with vorticity $m_a$ in replica $a$ as $\Pi_{ab}m_am_b A_c$ with $A_c\approx \sqrt{J/\gamma}$.
Adding this core energy, we can then expand the partition function in a Coulomb-gas like expansion
\begin{align}
Z\approx \int D \vartheta e^{-\int \frac{1}{8\pi^2\rho_s}\(\Pi\d_\mu\vartheta\)^2}\sum_{\{n^v_a\}}\[e^{-A_c \sum \Pi_{ab} n^v_an^v_b} e^{i\sum n^v_a\Pi_{ab}\vartheta_b}\].
\end{align}
Keeping only the minimal inter-replica vortex terms where $n_a=1$, $n_{b\neq a}=-1$, and $n_{c\neq a,b}=0$ in the sum (or more generally performing Poisson summation), gives  the dual Lagrangian of the main text with vortex fugacity: $\lambda\approx e^{-2A_c}$ where the factor of $2$ comes from the composite nature of the inter-replica vortices.

The relevance of the cosine term is determined by the scaling dimension of the vertex operator $e^{i(\vartheta_a-\vartheta_b)}$. Note that the vertex operator is unchanged if we replace $\vartheta_{a,b}\rightarrow (\Pi\vartheta)_{a,b}$ since this amounts to adding and subtracting the same term in the exponent. Hence, we can write correlators of the composite vertex operator as just the square of the correlators of a single-species:
\begin{align}
\<e^{i(\vartheta_a(r)-\vartheta_b(r))}e^{-i(\vartheta_a(0)-\vartheta_b(0))}\>_{g=0} \approx \frac{1}{r^{4\pi\rho_s}} \equiv \frac{1}{r^{2\Delta_2}},
\end{align}
where $\Delta_2$ is the scaling dimension of the composite inter-replica vortices (i.e. with absolute vorticity two compared to a single-replica vortex). These terms become relevant if $\Delta\leq 2$, identifying the KT transition at:
\begin{align}
\(\rho_s\)_\# = \frac{1}{\pi},
\end{align}
which, as noted, is half the usual value at an ordinary KT transition, which reflects that the minimal single-replica vortices are excluded.

\subsection{Steady-State observables and Boundary Criticality}
Observables in the replica field theory consist of products of terms like $\<1|O|\rho\>$ where $O$ is some $\sigma^z$-diagonal operator. For qubits (spins-1/2), the left boundary vector, $\<1| = \otimes_i \frac12 \(\<\up_i| +\<\down_i|\)$ in the $\sigma^z$ basis, i.e. corresponds to the $+x$ polarized product state of the spins, which we will continue to use for general spin size $S$. In the coherent state language this corresponds to $\(\theta,\phi\)=\(\frac{\pi}{2},0\)$ for all $x$ at the final time, $t$, which sets boundary conditions $\phi=0$.

To describe the steady state, we take $t=0$, and extend the imaginary time-evolution evolution back to $-\infty$, so that the path integral is defined on the half-plane $t\in (-\infty,0]$, $x\in\mathbb{R}$.  Equal-time observables thus correspond to boundary observables at $t=0$ in the field-theory. For future reference, we note that the boundary conditions $\phi_a(t=0,x)=0$ imply boundary conditions $\d_t\vartheta_a=0$ for the dual field (since $0=\d_x\phi =\d_t\vartheta/2\pi$).

To compute boundary correlators, it is useful to decompose the field $\phi(t,x)$ into left- and right- movers (a.k.a. holomorphic and anti-holomorphic parts): $\phi(z,\bar{z}) = \phi(z)+\bar\phi(\bar{z})$ with complex coordinates $z=x+it$, with dual field $\vartheta(z,\bar{z}) = \phi(z)-\bar\phi(\bar z)$. We see that $\phi(t=0,x)=0$ implies $\phi(z) =- \bar\phi(\bar{z})$ at $t=0$, i.e. $\vartheta(z,\bar{z})|_{t=0} = 2\vartheta(z)|_{t=0}$. Therefore, the boundary scaling dimension of $\vartheta$ \emph{doubles} compared to its bulk value, since: 
$\<e^{i\vartheta(z,\bar{z})}e^{-i\vartheta(w,\bar{w})}\>|_\text{bulk} 
= |\<e^{2i\vartheta(z)}e^{-2i\vartheta(w)}\>| 
= |x|^{-4\Delta_{\vartheta,\text{bulk}}}$ 
at $t=0$.

It is convenient to compute correlators in the dual sine-Gordon action. In the IRSF (charge-fuzzy) phase, where cosine terms are irrelevant, the boundary correlators of $\vartheta$ decay like:
\begin{align}
\<:\Pi\vartheta_a(t,x)\Pi\vartheta_b(0,0):\> = -\Pi_{a,b} 4\pi \rho_s \log (|x|/a),
\label{eq:varthetacorr}
\end{align}
where $::$ denotes normal ordering, $a$ is a UV cutoff (lattice spacing), and this expression is valid when $t\ll x$ so that both operators are near the $t=0$ boundary

\paragraph{Charge-correlators}
The charge is defined by $\sigma^z_a = \frac{1}{\pi}\d_x \vartheta_a$. Inserting Eq.~\ref{eq:varthetacorr} above and taking the replica limit $Q\rightarrow 1$, this gives quadratic decay of charge-fluctuations in the IRSF (charge-fuzzy) phase:
\begin{align}
\E\[\<\sigma^z_x\sigma^z_0\>-\<\sigma^z_x\>\<\sigma^z_0\>\] &= 
\lim_{Q\rightarrow 1} \frac{1}{Q(Q-1)}\Pi_{ab}\<\sigma^z_a \sigma^z_b\>_\text{replica-theory}
%\nonumber\\&
=-\(\frac{1}{\pi}\d_x\)^2 \[4\pi\rho_s\log (|x|/a) \]=\frac{4\rho_s}{\pi}\(\frac{a}{x}\)^2.
\end{align}

In the gapped phase (charge-sharp) phase, the inter-replica fluctuations of $\vartheta$ are pinned by the relevant cosine terms and acquire mass $\sim \lambda$, so that correlations decay exponentially with characteristic length $\xi \sim \sqrt{\gamma/J}e^{-2\sqrt{\gamma/J}}$.

\paragraph{String-correlators}
Next, we consider the dual vortex correlations, that are probed via those of string operators: $W_{I,a} = \prod_{x\in I}e^{-i\pi \sum_{x\in I} \sigma^z_{a,x}/2}$ defined on interval $I=[x,x']$. In the field-theory, $\sigma^z = \d_t\phi= \frac{1}{\pi}\d_x\vartheta$, so this operator is: 
$W_{[x,x'],a} = e^{-i\frac\pi 2\int_{x'}^x \frac{1}{\pi}\d_{y}\vartheta_a(y)dy}=e^{-i\vartheta_a(x)/2}e^{i\vartheta_a(x')/2}$,
which inserts (removes) a 1/2-vortex from $x$ ($x'$) respectively. Second moments of this correlator such as:
\begin{align}
\mathbb{E}_{u,m}\[\<W_{I}\>\<W_I\>\] &\rightarrow \<e^{-i(\vartheta_a(x)-\vartheta_a(x')+\vartheta_b(x)-\vartheta_b(x'))/2}\>|_{t=0}
= 1/|x|^{2\pi \rho_s} ~~~~\(\underset{\text{at $p_\#$}}{=} \frac{1}{|x|^2}\).
\end{align}
We note that, as a dual ``disorder parameter", the $W$ correlator decays becomes slower as measurement strength is increased towards $p_\#$. For $p>p_\#$, where vortices become relevant, $W$ becomes truly long-range ordered, and the exponent for its power-law decay drops discontinuously to zero in the thermodynamic limit. We note that, it is typical for finite-size numerics on KT transitions to show a rather broad regime of power-law like delay even in the gapped phase, and that the most reliable means of identifying the thermodynamic limit tends to be based on locating where the power-law decay of these type of observables achieves its KT (or in this case modified-KT) value.

\paragraph{Charge-variance}
Another closely related quantity is the variance of charge in an interval of size $x$, which in the IRSF phase grows logarithmically with $x$:
\begin{align}
\text{Var}_q(x) = \sum_{0<i,j<x}\E\[\<\sigma^z_i\sigma^z_j\>_c\] \approx  \left\<\(\int_0^x \frac{d\vartheta}{\pi}\)^2\right\> = \frac{8\rho_s}{\pi} \log |x|/a +\dots,
\end{align}
where $\<\cdots\>_c$ refers to the connected part of the correlator, and the $\dots$ reflect non-universal subleading terms. The coefficient of the logarithm is universal in the charge-fuzzy/IRSF phase, and decreases monotonically with measurement strength to achieve a minimal value of $8/\pi^2\approx 0.8$ at the charge-sharpening transition. In the charge-sharp phase, the logarithmic growth of variance saturates to a constant plateau at distance scales beyond the correlation length $\xi \sim \exp\[\text{const.}/\sqrt{|p-p_\#|}\]$. 

These predictions are consistent with the TEBD data, which show logarithmic growth of the charge variance with interval size $\ell$. Fitting the coefficient of the log, we find that it decreases to the expected critical value $\approx 0.8$ at the same value of $p_\#$ identified from the dual order parameter. This agreement is obvious in the field theory: the dual order parameter is simply the exponential of the charge variance, and since any Gaussian random variable $X$ satisfies $\<e^{ i aX}\> = e^{-\frac{a^2}{2}\<X^2\>}$. Nevertheless, observing this consistency in the discrete circuit model numerics provides a highly non-trivial check that the charge sharpening transition is described by the predicted free-boson CFT.

\section{Hydrodynamics of charge sharpening for $p \ll p_c$}

In this section we provide a heuristic but model-agnostic derivation of our central result that the fuzzy phase has algebraic charge correlations. This derivation only applies for $p \ll p_c$ but does not depend on the large-$d$ limit. The strategy we will adopt is to compute the equal-time, finite-wavevector correlator:
\begin{align} 
C_n(k) = \int dx  \left[ \langle n(x) n(0) \rangle - \langle n(x) \rangle \langle n(0) \rangle \right] e^{i kx}.
\end{align}
 $C_n(k)$ is heuristically the steady state variance per length of the charge in a block of size $2\pi/k$. 

First consider the case of $p = 0$. Here, the steady state is the infinite temperature state, where charge is uncorrelated from site to site; therefore $C_n(k)$ is $k$-independent at small $k$. For diffusive dynamics, this featureless steady state of $C_n(k)$ is achieved by balancing two effects: diffusion, which causes $C_n(k)$ to decay as charge fluctuations spread across the system, and noise, which causes it to increase. These effects are captured by the noisy diffusion equation
\begin{equation}
\partial_t C_n(k) = B k^2 - D k^2 C_n(k),
\end{equation}
where $D$ is the diffusion constant and the coefficient $B$ (related to $D$ by the fluctuation-dissipation theorem) depends on the details of the model. 

For $p > 0$, measurements in a region of size $L$ sharpen the charge on a timescale $\sim L/p$. The corresponding rate scales (in this naive estimate) as $k$, and thus dominates over diffusion. Thus, for $p > 0$ the steady state is achieved by balancing the increase in charge uncertainty due to charge spreading (i.e., ``noise'') against the decrease due to sharpening. 

We now estimate the rate at which measurements sharpen charge. We will find that this sharpening rate is itself proportional to the charge variance---a result that is intuitive, since the more uncertain the charge is the more one learns from measurements. For simplicity, consider a chain of qubits of length $L$, with an initial state that is a mixture of two charge states, $N$ and $N + 1$, with probabilities $\mathfrak{p}$ and $1-\mathfrak{p}$ respectively. We assume that the state gets scrambled between measurements so that, on a given measurement of the charge at a particular site, the probability of measuring the site to be occupied is $N/L$ if the charge state is $N$, and $(N+1)/L$ otherwise. For this model, the initial charge variance is $\mathfrak{p}(1-\mathfrak{p})$. The final charge variance is the variance conditional on each measurement outcome, weighted by the Born probability of that outcome. A straightforward calculation shows that, for large $L$, the variance is reduced by an amount $[\mathfrak{p}(1-\mathfrak{p})]^2/(L(N-L))$ per measurement. Therefore, the change in variance is proportional to the square of the initial variance, as claimed. This observation generalizes to arbitrary discrete distributions, though it becomes increasingly tedious to compute this rate. 

Using this estimate of the sharpening rate, we get that
\begin{equation}
\partial_t C_n(k) = B k^2 - \kappa p C_n(k)^2 - D k^2 C_n(k).
\end{equation}
The steady state at small $k$ is achieved by balancing the first two terms on the rhs of this equation, yielding the result: 
\begin{align}
C_n(k)\vert_{\text{steady-state}} \sim |k|/\sqrt{p},
\end{align}
in agreement with the field theory prediction.

\section{Modified percolation for charge degree of freedoms}

\begin{figure}[b!]
\includegraphics[width=0.65\textwidth]{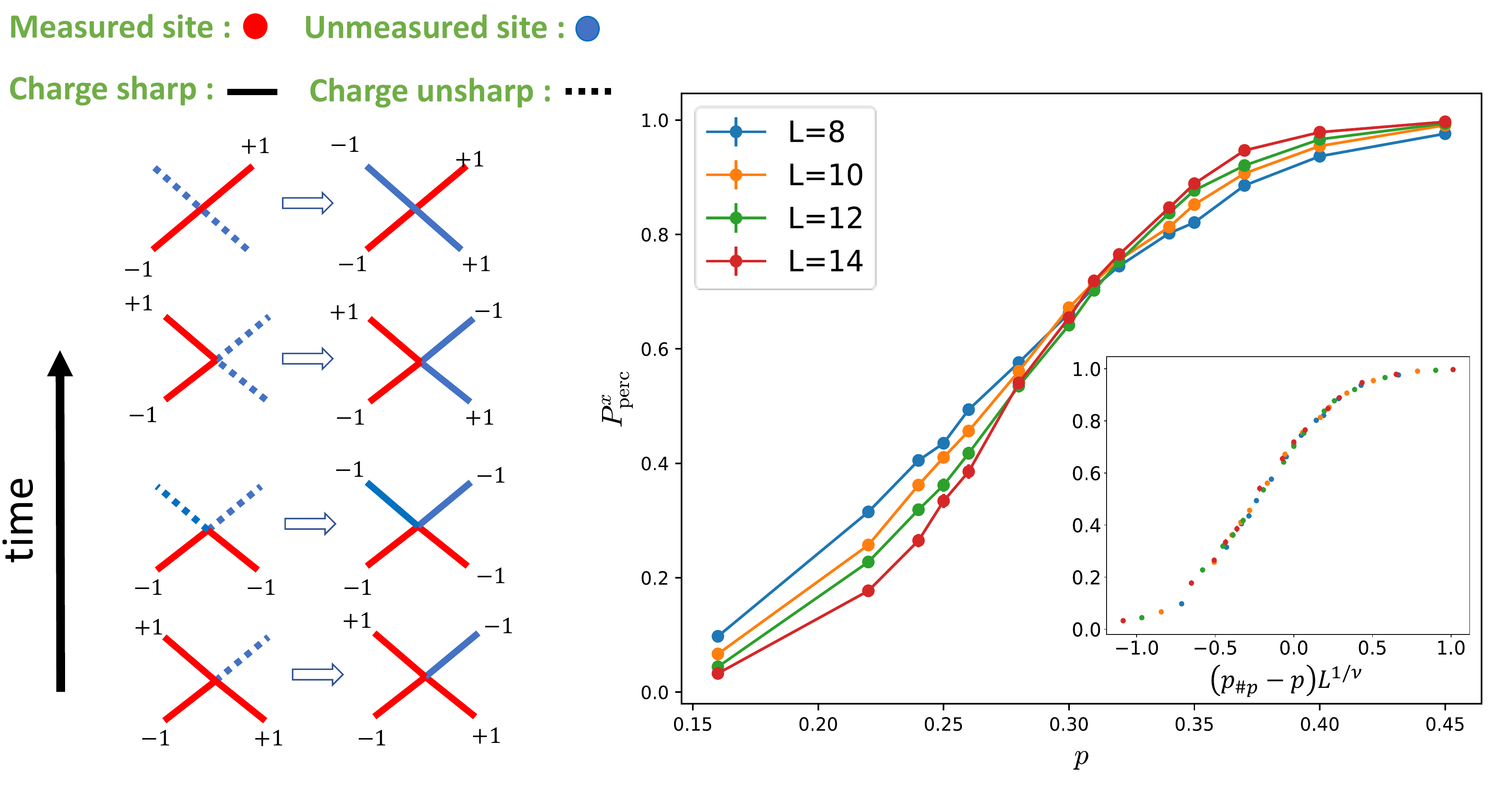}
\caption{\textit{Left.} Illustration of some ways in which unmeasured sites become charge sharp. Other scenarios can be obtained similarly.
%are obtained via transformation $-1\leftrightarrow +1$, and spatial reflection. 
The bold links percolate at around $p_{\#p}\approx 0.31$ and red links percolate at $p_c=0.5$. \textit{Right.} Probability for a cluster of charge sharp sites to wrap around the spatial direction vs $p$. The inset shows collapse with $p_{\#p}=0.31$ and $\nu=4/3$.}\label{Fig: percolation figure}
\end{figure}

In this section we argue that for large Hilbert space dimension for the qudits, the sharpening transition must occur inside the volume law phase in any spatial dimension $d$. The basic idea is that charge conservation induces correlations between measurement outcomes, and allows multiple measurements to extract more information about charge than they could about neutral degrees of freedom. 
A measured site is charge sharp in the sense that the projective measurements will collapse the charge to either $-1$ or $+1$. In addition to this, charge conservation can dictate the charge of an unmeasured site based on the outcome of nearby measurements. For example, measuring three out of four legs of a gate determines the charge at the fourth. Figure \ref{Fig: percolation figure} shows various related scenarios where unmeasured sites become sharp. Thus we expect charge sharp sites to start percolating at a smaller value $p_{\#p}$ compared to that for the percolation of measured sites; that is $p_{\#p}<p_c$ where $p_c$ is the percolation transition of the measured links. In the limit we are working in in this paper (Hilbert space dimension $\rightarrow \infty$), the entanglement transition coincides with the percolation transition for the measured links. Thus, the percolation of the charge sharp sites happens inside the volume law phase.
The presence of the percolation transition for charge sharp sites implies that a sharpening transition must happen at value $p_\#\leq p_{\#p}$. In other words, the sharpening transition must happen inside the volume law phase. Note that this result is true for any dimension. Thus, we have the result that the sharpening transition, in all dimensions, must occur \textit{inside} the volume law phase, well separated from the entanglement transition.

To numerically study this modified percolation in $1+1$ dimension, we entangle the charge at every space-time point to a ancilla qubit. A site at $(x,t)$ is then charge sharp iff the corresponding ancilla at $(x,t)$ becomes charge sharp. (Note that the ancilla becoming charge sharp is a stronger condition than the ancilla getting disentangled from the system. In general the ancilla might become disentangled without becoming sharp.) This allows us to determine space time points with sharp charge.  A standard percolation analysis then shows that the sharp sites start percolating at $p_{\#p}\approx 0.31$ which is much less than the percolation transition for the measured sites at $p_c=0.5$. Figure \ref{Fig: percolation figure} shows the probability for sharp sites to percolate along spatial direction. These curves show a clear crossing with increasing system size at $p_{\#p}$, and collapse upon rescaling with the standard percolation form with correlation exponent $\nu=4/3$.

We emphasize that this percolation of charge-sharp sites does not reflect the true charge sharpening transition that occurs at a smaller measurement value, $p_\#\approx 0.2$, and has $\nu=\infty$ KT-like scaling rather than percolation scaling.  In fact we conjecture that this sharp-site percolation transition may not be visible in any physical degrees of freedom for generic models where measurements are not perfectly projective (as this blurs the distinction between sharp and unsharp sites). However, the sharp-site percolation threshold clearly upper-bounds the critical measurement probability for sharpening: $p_c>p_{\#p}\geq p_\#$ in the projective measurement limit, supports the argument that the charge-sharpening transition generically occurs in the volume-law entangled phase. We also note that the charge-sharpening transition identified as a finite-size crossing in the fraction, $N_0$ of exactly-charge-sharp trajectories in Ref.~\cite{agrawal2021entanglement} occurs at $p\approx p_{\#p}$, and may be probing this auxiliary critical point rather than the true charge-sharpening transition at $p_\#\approx 0.2$ (an issue exacerbated by the strong finite-size corrections to scaling near KT transitions).

\end{document}